\documentclass[10pt,twocolumn]{IEEEtran}
\usepackage{graphicx}          
\usepackage{amsmath}
\usepackage{tabularx}
\usepackage{amsfonts}
\usepackage{url}
\usepackage{verbatim}
\usepackage{times,epsfig}
\usepackage{makecell}
\usepackage{psfrag}
\usepackage{subfigure}
\usepackage{stfloats}
\usepackage{setspace}
\usepackage{amssymb}
\usepackage{multirow}
\usepackage{colortbl}
\usepackage[justification=centering]{caption}
\usepackage{fancyhdr}
\usepackage[table,xcdraw]{xcolor}

\begin{document}
	
	\pagenumbering{arabic}
	
		\title{Secure and Reliable Transfer Learning Framework for 6G-enabled Internet of Vehicles}
	\author{Minrui Xu, Dinh Thai Hoang, Jiawen Kang*, Dusit Niyato, \emph{Fellow, IEEE}, Qiang Yan, Dong In Kim, \emph{Fellow, IEEE}
		
		\IEEEcompsocitemizethanks{
		Minrui Xu, Jiawen Kang, Dusit Niyato are with School of Computer Science and Engineering, Nanyang Technological University, Singapore. 
		Dinh Thai Hoang is with School of Electrical and Data Engineering, University of Technology Sydney, Australia. 
		Qiang Yan is with WeBank Co., Ltd., China. Dong In Kim is with the College of Information \& Communication Engineering, Sungkyunkwan University, South Korea. (\textit{*Corresponding author: Jiawen Kang})
		}
	}
	\maketitle
	\pagestyle{headings}

	\begin{abstract}
	
    In the coming 6G era, Internet of Vehicles (IoV) has been evolving towards 6G-enabled IoV with super-high data rate, seamless networking coverage, and ubiquitous intelligence by Artificial Intelligence (AI). Transfer Learning (TL) has great potential to empower promising 6G-enabled IoV, such as smart driving assistance, with its outstanding features including enhancing quality and quantity of training data, speeding up learning processes and reducing computing demands. Although TL had been widely adopted in wireless applications (e.g., spectrum management and caching), its reliability and security in 6G-enabled IoV were still not well investigated. For instance, malicious vehicles in source domains may transfer and share untrustworthy models (i.e., knowledge) about connection availability to target domains, thus adversely affecting the performance of learning processes. Therefore, it is important to select and also incentivize trustworthy vehicles to participate in TL. In this article, we first introduce the integration of TL and 6G-enbaled IoV and provide TL applications for 6G-enabled IoV. We then design a secure and reliable transfer learning framework by using reputation to evaluate the reliability of pre-trained models and utilizing the consortium blockchain to achieve secure and efficient decentralized reputation management. Moreover, a deep learning-based auction scheme for the TL model market is designed to motivate high-reputation vehicles to participate in model sharing. Finally, the simulation results demonstrate that the proposed framework is secure and reliable with well-design incentives for TL in 6G-enabled IoV.

	\end{abstract}

	
	\section{Introduction}
	\bibliographystyle{ieeetr}
	
    The unprecedented evolution of wireless communication technologies, e.g., from 4G to 5G and beyond, has paved the way for enormous advanced vehicular network applications. Particularly, 6G technologies can provide seamless and ubiquitous communications for large-scale and ad-hoc vehicular networks~\cite{ji2020survey}. Therefore, there is a strong push for traditional Internet of Vehicles (IoV) to evolve into 6G-enabled IoV to meet stringent performance requirements. In detail, 6G technologies are expected to provide future mobile applications in 6G-enabled IoV with unprecedented benefits~\cite{saad2019vision}, such as super-high data rates (e.g., up to 1Tb/s), very broad frequency bands (e.g., 73GHz-140GHz and 1THz-3THz), and less than 1-millisecond end-to-end latency.
    
    In the near future, 6G-enabled IoV is expected to facilitate a safer, more efficient, and intelligent traffic system, which strictly requires heterogeneous communications, latency-critical applications, and scalable vehicular networks. To meet these requirements, Artificial Intelligence (AI) has been regarded as a core component to empower 6G-enabled IoV~\cite{ji2020survey}. Machine Learning (ML), e.g., deep learning (DL) and reinforcement learning can be implemented to design and optimize IoV architectures and network orchestrations in a cost-efficient manner. In addition, through a huge amount of data collected from entities in IoV (e.g., vehicles and infrastructures), AI-based solutions are able to synthesize, analyze, and provide decisions for optimal operations, thereby meeting stringent security, high-mobility, and efficiency requirements of 6G-enabled IoV. For example, AI can assist drivers by recognizing, searching, and sharing local and roadside information, e.g., traffic signs and adaptive speed signs~\cite{zhou2021two}. This smart driving assistance thus can significantly improve driving safety, reduce energy consumption, and enhance traffic management efficiency in 6G-enabled IoV.

	Although being regarded as an integral component of 6G-enabled IoV, ML has been facing challenges for practical implementation in the IoV environment due to  high mobility, dynamics, and heterogeneity of vehicular networks~\cite{nguyen2021transfer}. In particular, the deployment of ML in 6G-enabled IoV is constrained by dynamic wireless environments, lack of labeled data, long training process, and limited computation capacity of vehicles. Fortunately, Transfer Learning (TL)~\cite{yang2020transfer} has been recently introduced with many applications in wireless networks, e.g., spectrum management and signal recognition. TL can help to develop intelligent solutions to maximize data utilization efficiency from the same or similar domains, thereby remarkably enhancing AI-based solutions for IoV \cite{nguyen2021transfer}. As a result, the convergence of TL and 6G-enabled IoV is expected to significantly speed up the learning process and reduce computation costs, thus meeting the extremely high requirements of 6G-enabled IoV.

 
 However, due to the utilization of knowledge from different sources, which might be untrusted, to improve learning efficiency, the reliability and security of TL are the main concerns for its deployment. Thus, in this article, we introduce a novel framework that provides secure and reliable services for deploying TL in future 6G-enabled IoV. Specifically, we introduce the concept of reputation as a metric to evaluate the reliability and trustworthiness of pre-trained models built and owned by vehicles, and thus design a reputation-based trainer selection scheme for FL. To achieve secure and efficient decentralized reputation management, consortium blockchain is leveraged for reputation management. Finally, to motivate high-reputation vehicles participating in TL model transferring, a deep learning-based auction (DLA) scheme is designed to incentivize trustworthy vehicles. 
	
    The main contributions of this article are summarized as follows:
      \begin{itemize}

        \item We first introduce TL, 6G, and IoV together with their integration, which envisions a blueprint for future intelligent transportation systems. To enable efficient TL deployment, we design a novel model trading market for 6G-enabled IoV. The proposed model trading can effectively improve the quality of ML in TL environments by encouraging good source model owners to participate and contribute to the model transferring.
        \item We employ reputation as a reliable metric to select trustworthy pre-trained models for reliable TL. We then design a distributed reputation calculation and management scheme by utilizing blockchain technology to regulate the reputation in a decentralized and secure manner. With the proposed metric and scheme, TL can be performed in IoV systems in a more secure and reliable way.
        \item To achieve dynamic and reliable model trading, we propose the DLA scheme jointly considering the reputation of model sellers and bids of model buyers. This scheme is effective to maximize revenue for sellers, i.e., source model owners, while guaranteeing individual rationality and incentive compatibility for buyers, i.e., target model owners.
        Simulation results show that the proposed TL framework can assure secure and reliable services as well as satisfactory incentives for vehicles.
        
        \end{itemize}
	

	\section{Fundamentals of Transfer Learning}
	\label{sec:Overview}
	\begin{figure*}[!]
		\centering
		\includegraphics[width=0.7\linewidth]{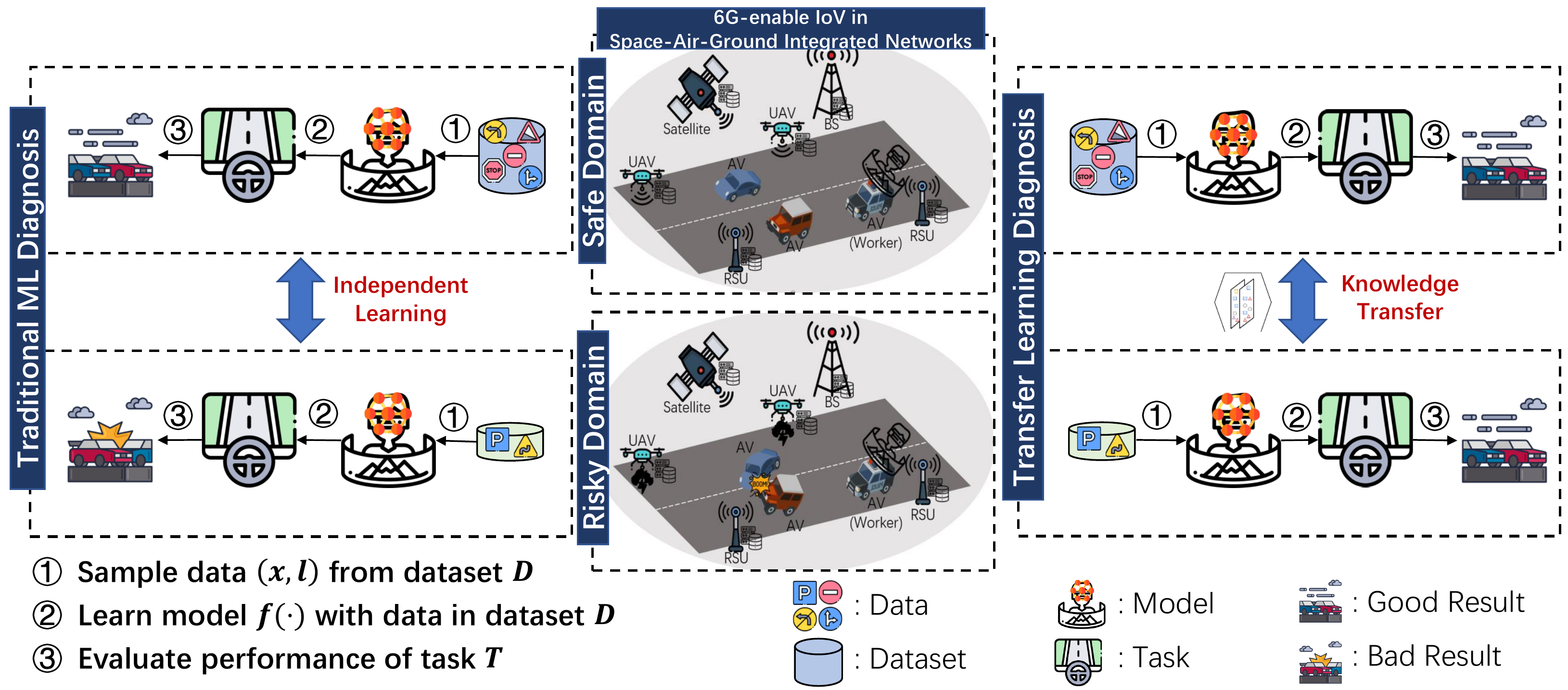}
		\caption{Tradition machine learning vs. Transfer Learning for 6G-enabled IoV in Space-Air-Ground Integrated Networks.}
		\label{fig_ML_vs_TL}
	\end{figure*}
	
	
	\subsection{Overview}
	Transfer learning has recently been introduced as a highly effective machine learning solution that can address many existing problems of traditional ML methods~\cite{yang2020transfer}. TL can effectively extract valuable knowledge from learning tasks in source domains to improve the learning performance in similar target domains. In Fig. 1, we provide an example of using traditional ML and TL for 6G-enable IoV in Space-Air-Ground Integrated Networks (SAGINs)~[R7], which aim to provide ubiquitous connectivity by utilizing space, air, and ground vehicles and infrastructures. In SAGIN, AI can be adopted for optimal path planning of satellite networks to improve the driving safety of ground vehicles. However, the performance of traditional ML highly relies on the collected data, which can be costly and unbalanced. In detail, data from the safe domain, e.g., a road segment with rare accidents, is rich while data from the risky domain, e.g., a junction with frequent accidents, is scarce. This leads to unsatisfactory performance of ML models in risky tasks, such as car crashes. Fortunately, TL can address the data unbalance issue over traditional ML techniques by leveraging valuable knowledge from similar environments or applications. Therefore, TL can overcome the lack of labeled data in risky tasks. Additionally, the shared knowledge, along with knowledge collected and extracted from previous experiences, can also be utilized to reduce computation workloads required for model training as well as to speed up slow learning processes of traditional ML techniques. Furthermore, the knowledge from other sources can be transferred in the form of weights, such as learning rate and random seed. As a result, TL does not incur much communication overhead, and data privacy is also protected. With such outstanding benefits, TL has been considered to be an indispensable part of future wireless networks~\cite{nguyen2021transfer}.
	
	
    
	\subsection{Basic Concepts}
	\begin{itemize}
		\item \textit{Domain:} A domain is defined by a feature space and a marginal probability distribution. The marginal probability distribution is the probability distribution of different types of signal in the dataset.
		\item \textit{Task:} Given a domain, a task can be defined by a label space (e.g., a set of all driving decisions) and a predictive function. By training with datasets of the source and the target domains, the predictive function can predict labels for given data samples. 
	\end{itemize} 
	
	Based on the domain and task, TL can be defined as follow: given a source domain with a corresponding source task and a target domain with a corresponding target task, the goal of TL is to learn the target predictive function by leveraging the knowledge gained from the source domain and the source task. In TL, either the source domain and the target domain are different or the tasks of both domains are different. Take autonomous driving as an example, this means that, either the image features of the set of source driving experience and the set of target driving experience are different (i.e., driving under different scenarios), or the marginal distributions of the two sets are diverse. Therefore, TL can be applied in various applications in 6G-enabled IoV.
	
	\subsection{Classification of Transfer Learning}
	
	
	Based on the similarity between domains and tasks as well as the availability of labeled data, TL can be classified into: 
	\begin{itemize}
		\item {\textit{Inductive TL}}: The source and target domains are the same, but the source and target tasks are different. For inductive TL, the target data should be labeled. Depending on the availability of labeled data at the source domain, inductive TL can be further classified into \textit{self-taught learning} and \textit{multi-task learning}. In \textit{self-taught learning}, the source data is not labeled, so they can only be leveraged to reduce the feature space in the target domain. In contrast, labeled source data is available in \textit{multi-task learning}. Thus, the source model's weights and parameters can be transferred to the target model.
		
		\item \textit{Transductive TL}: The source and target domains are different, but the source and target tasks are the same. For transductive TL, labeled target data is not required, while the source data must be labeled. 
		
		{\item \textit{Unsupervised TL}}: The target and source tasks are different and there is no labeled data in both domains. Commonly, the knowledge gained from the unsupervised learning process at the source task can be transferred to serve as the initial point for learning the target tasks.
	\end{itemize}
	
	The choice among these three TL classifications is determined by the availability of training data in learning problems. When labeled data is available in target domains, inductive TL can be utilized to accelerate the learning process of TL. If labeled data is unavailable in target domains but available in source domains, transductive TL is effective to be leveraged to enhance quality and quantity of data in target domains. Finally, when labeled data is unavailable in both source and target domains, unsupervised TL can be used to transfer the knowledge between source and target domains.
	
	\subsection{Strategies of Deep Transfer Learning}
	Imitating the brain structure of creatures, DL uses a multi-layered architecture called Deep Neural Networks (DNN). DNN is trained to perform specific tasks such as classification, clustering, or regression. With the training data, DNN updates its knowledge, which is represented by its parameters. Using the knowledge acquired during its learning phase, DNN executes the learned task. A trained DL model can be regarded as knowledge obtained from training data, including its architecture and parameters. Therefore, DL provides an effective way to transfer knowledge from one domain to another one, namely,  Deep Transfer Learning (DTL)~\cite{yang2020transfer}. There are four typical strategies used in DTL:
	
	\begin{itemize} 
		\item \textit{Off-the-shelf pre-trained models}: DL model training is a data-hungry and time-consuming work. Fortunately, such an inefficient training process is relieved by directly leveraging pre-trained models, which are trained from neighboring domains, for target tasks. For example, in traffic sign classification tasks, the pre-trained models, such as VGG16 (storage size: 512 MB), GoogLeNet (40 MB), and ResNet50 (97 MB), can be easily obtained from common DL libraries.
		\item \textit{Pre-trained models as feature extractors}: Extracting features is an important procedure in ML algorithms, which directly affects decision-making for these algorithms. In DL, DNN can automatically learn the features extraction from the training data.
        Therefore, knowledge from the source domain is embedded in this new representation, which improves the learning process in the target domain.
		\item \textit{Fine-tuning pre-trained models}: Instead of leveraging all parameters of pre-trained models directly, certain parts or a whole pre-trained source model can be continuously fine-tuned with target data to further improve the performance of the TL model further. The fine-tuned of pre-trained models can be performed in Weight Initialized and Selective Fine-tuning. 
		\item \textit{Domain adaptation}: Adapting a domain refers to transferring knowledge from one or more source domains and improving the performance of the target learner. Domain adaptation is often used in TL to reduce the differences between domains.
	\end{itemize}
	
	However, the best strategy selection for DTL is not straightforward, which should consider multiple factors including the size of target data and similarity between the source and target domains. For example, when target labeled data is large, Weight Initialization is an effective solution since overfitting is not a considerable concern. In contrast, when target data is small, we can use a pre-trained model as a feature extractor. Moreover, if source and target data are similar, we can use the whole pre-trained model. Otherwise, it is better to transfer only some general features from the first few layers of DNN.

		\subsection{Potential Applications of TL in 6G-enabled IoV}
\begin{table*}[]
\label{table:table1}
\caption{Potential Applications of Transfer Learning in 6G-enabled IoV}
\centering\small\begin{tabular}{|c|c|c|c|}
\hline
\rowcolor[HTML]{C0C0C0} 
{\color[HTML]{333333} 6G-enabled IoV Applications} & Typical Services                                        & TL Strategies                 & TL Advantages                                                      \\ \hline
Mobile Extended Reality                           & \makecell[c]{Minimizing breaks in presence \\ of XR users~\cite{chen2019federated}}                   &  \makecell[c]{Fine-tuning \\ pre-trained models}           &  \makecell[c]{Enhancing quality of experiences,\\ reducing computing demands}    \\ \hline
Mobile Digital Twins                              & \makecell[c]{Mitigating unreliability in \\ long-distance communication~\cite{lu2020low}} &  \makecell[c]{Pre-trained models \\ as feature extractors} &  \makecell[c]{Improving driving safety,\\ protecting drivers' privacy}               \\ \hline
\makecell[c]{UAV-assisted \\Autonomous Driving}                   &  \makecell[c]{Providing ultra-high reliability \\ and connectivity~\cite{hieu2021transferable}}      &  \makecell[c]{Off-the-shelf \\pre-trained models}         &  \makecell[c]{Addressing lack of labeled data, \\speeding up learning processes} \\ \hline
\end{tabular}
	\vspace*{-4mm}
\end{table*}

	\subsubsection{Mobile Extended Reality}
	Mobile Extended Reality (XR) refers to various technologies, such as augmented, mixed, and virtual reality employed to digitally enhance vehicular environments and human interactions in real-time. The rapid development of 6G networking technologies and wearable devices is paving the way for XR. However, this also brings forth several challenges. Particularly, drivers' demands for more high-quality and personalized experiences have been rapidly increasing in 6G-enabled IoV. For example, a federated learning algorithm in~\cite{chen2019federated} was used to proactively determine vehicles' orientations and mobility to minimize breaks in presence that can detach the XR users from their virtual world. However, conventional ML techniques might not be effective as they may require a lot of data and training time. By utilizing TL, data collected from vehicles' individual interactions can be collected to train a generalized ML model, and then the model is transferred to other nodes (e.g., vehicles) for fine-tuning, thereby enhancing the quality of experiences and reducing computing demands simultaneously.
	
\subsubsection{Mobile Digital Twins}
Digital twins utilize the digital monitoring of physical vehicles to achieve real-time and accurate operations for 6G-enabled IoV. With the availability of a massive amount of data, ML is an essential solution for digital twins to leverage such data to enable various cutting-edge vehicular applications. For instance, a federated learning-empowered digital twin wireless network was introduced in~\cite{lu2020low} to mitigate the unreliability in long-distance communication between vehicles and infrastructures. However, due to strict requirements of driver-privacy protection and data sharing, the ML-based driving virtual assistants usually do not have sufficient data for timely traffic analysis and monitoring. To solve this problem, TL can be used to leverage knowledge from similar experiences (i.e., source domain) to each vehicle (i.e., target domain) aiming to improve the ML-based personal driving assistant's performance for driving safety.

	
	\subsubsection{UAV-assisted Autonomous Driving}
	
	Despite its rapid development, autonomous driving is still facing challenges in 6G-enabled IoV. Particularly, Unmanned aerial vehicles (UAV)-assisted automated driving systems are network-dependent, i.e., they require continuous communication to function properly. In this context, ML can enhance these systems with ultra-high reliability and connectivity~\cite{ji2020survey}. TL can be applied into these systems which are impeded by lack of labeled data, to enhance the performance of the conventional ML techniques. Specifically, there are many factors affecting vision-based data, e.g., weather conditions and broken infrastructures. These variances are not always presented in a dataset collected for a specific area. Therefore, knowledge about these variances can be transferred from a more comprehensive dataset of different areas to improve the target models' robustness if such events occur. For example, to optimize the performance of the autonomous driving systems under environments with different objectives, uncertainties, and dynamics, a TL-based framework was developed in~\cite{hieu2021transferable} for enhancing scalability and reliability for joint radar and data communication systems. Thus, TL allows vehicles to learn an optimal policy quickly when they travel to new environments.
	
	Finally, the summary of these aforementioned potential applications is provided in Table I.

	
	
	\section{Challenges of Implementing Transfer Learning in 6G-enabled IoV}
	\label{sec:Solutions}

	\subsection{Technical Challenges of TL Algorithms}

	\subsubsection{Determine the Source Task} TL requires in-depth and task-specific analysis in domains of source and target tasks to determine whether these tasks are similar. For example, to perform similar resource management tasks in IoV, the source task can be chosen based on physical distance from source to target vehicles~\cite{yang2020transfer}. The reason is that surrounding vehicular networks of source and target tasks are similar, and hence their experiences are likely to be related. Alternatively, TL can transfer knowledge of the same vehicular task (e.g., pedestrian detection task) to different driving environments. However, 6G-enabled IoV typically has very wide coverage~\cite{ji2020survey}, leading to a large number of choices for source domains for TL applications. Moreover, several 6G-enabled IoV services, e.g., satellite communications, are specific to certain areas and groups of vehicles. If source and target domains are not correlated, the use of TL in 6G-enabled IoV even makes target models perform worse than just training with only target data. 
	
	\subsubsection{Determine What to Transfer} Following the choice of the source task, the domain knowledge needs to be transferred to target tasks. Determining what to transfer is important for TL to enable vehicular networks to cover a wide range of scenarios. However, in some scenarios, e.g., car crashes which are rare, obtaining the labeled data for TL training in the real world is intractable and even impossible. As such, transductive TL is typically used to transfer some related knowledge to handle the tasks in such scenarios. For instance, in emergency rescue tasks, the rescue vehicles can be trained to save life in simulation systems and then be dispatched in real-world~\cite{ji2020survey}. To further reduce fine-tuning time, some TL approaches to transfer the entire source model, while others transfer only the first few layers of DNN. However, this question depends on IoV scenarios that affect the fine-tuning time of TL.


	\subsection{Reliability and Security of TL}
	
	\subsubsection{Decentralized Environments}
	Most of the aforementioned TL model sharing schemes are centralized, i.e., the transfer of knowledge is governed by a single entity, e.g., a network operator. However, in 6G-enabled IoV with a massive scale of interconnected vehicles and infrastrutures~\cite{ji2020survey}, decentralized TL model sharing is a more promising approach. Particularly, valuable knowledge can be obtained from external or third-party sources such as neighboring networks, controlled by different operators, or vehicles. However, for such decentralized approaches, security and privacy challenges become a primary concern, e.g., attackers, eavesdroppers, and malicious vehicles. For example,  a model used for accident prevention in 6G-enabled IoV should have a higher degree of accuracy than that of an ordinary vehicular networks because a single false negative event (e.g., failed to predict a potential accident) could have very serious consequences. In recent years, blockchain, which can enhance decentralized network's security and privacy, has been successfully applied in various wireless applications. Thus, utilizing blockchain for decentralized TL is a attractive research direction. For instance, the authors in \cite{9016456} proposed a blockchain-based framework for DTL. By using blockchain and smart contracts, the proposed framework allows vehicles to securely and reliably share their models, data, and resources. Moreover, smart contracts facilitate automatic payment processes, which paves the way for crowdsourcing-enabled TL.
	
	\subsubsection{Model Quality Evaluation}
	
	TL is still vulnerable to adversarial attacks for data owners (e.g., vehicles) in 6G-enabled IoV can intentionally or unintentionally mislead the source model during TL processes. For instance, in the poisoning attack, an attacker sends a malicious source model to manipulate the parameters of the source model, resulting in the failure of the model transfer~\cite{yang2020transfer}. Furthermore, dynamic vehicular networking environments may lead to some unintentional adverse behaviors of source model owners. In addition, data owners may unintentionally update poor-quality models due to high-speed mobility or energy constraints, thereby degrading the learning performance. Thus, TL should be able to evaluate the quality of transferred models accurately to avoid such intentionally or unintentionally unreliable model sharing.
	
	\subsubsection{Efficient Incentive Mechanism for Model Sharing}
	TL can improve the effectiveness of current ML algorithms by transferring the knowledge of the source domain to the target domain. However, to motivate high-quality source model trainers to share their models, an efficient incentive mechanism for a model sharing market is also needed. In this market, source model trainers (e.g., vehicles) as sellers spend their data and computing resources to improve the inference ability of pre-trained models in source domains. On the other hand, model buyers in the market are the receivers, whose goals are obtaining the utility-maximized and reliable target models to perform their local tasks (e.g., traffic sign recognition tasks). Naturally, participants in the TL market are self-interested and attempt to maximize their utilities. Model sellers intend to maximize their revenue from model trading, while model buyers aim to receive the best models for local tasks. Without proper pricing and allocation mechanisms, trustworthy model sellers may be reluctant to share their models with potential buyers. Moreover, in 6G-enabled IoV which requires ultra-high reliable and ultra-low latency services~\cite{ji2020survey}, incentive schemes are required to perform decisions in real-time under dynamic network environments. As a result, the incentive scheme designed for the TL market to meet these stringing requirements has still not well investigated.
	
	\section{A Secure and Reliable Framework of Transfer Learning For 6G-enabled IoV}
	\label{sec:framework}
	
    In this section, we propose a secure and reliable TL framework for 6G-enabled IoV to address the aforementioned problems. In particular, this framework includes three phases, i.e., blockchain-empowered decentralized model sharing, reputation for model quality evaluation, and learning-based auction for model trading. Thus, the proposed TL framework for 6G-enabled IoV can offer decentralized and real-time services.
	
	\subsection{Blockchain Empowered TL Model Trading Markets} 
\begin{figure*}[!]
	\vspace{-0.2cm}
		\centering
		\includegraphics[width=0.75\textwidth]{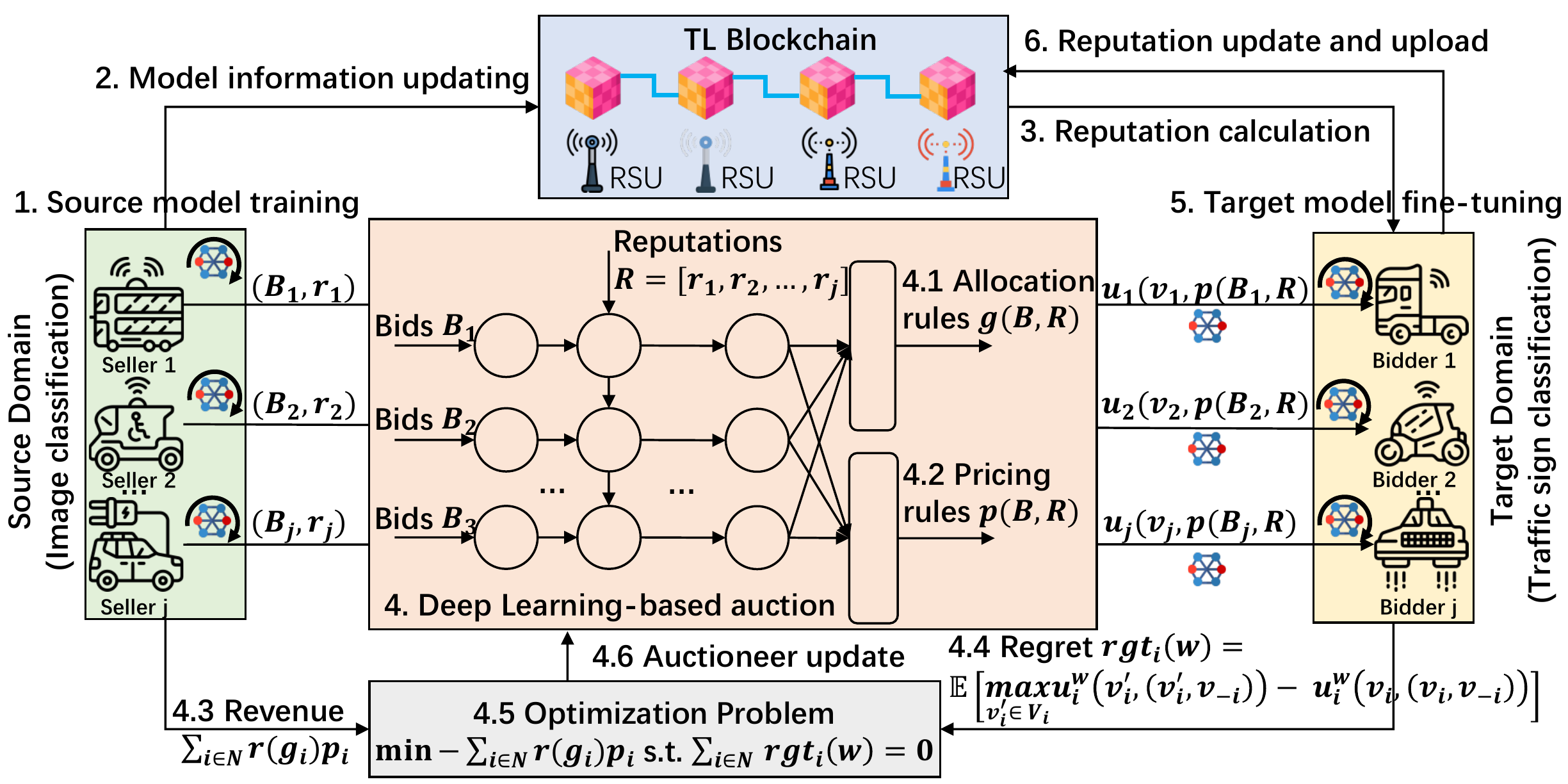}
		\caption{A secure and reliable TL framework for 6G-enabled IoV including three main phases, i.e., blockchain-empowered decentralized model sharing, reputation for model quality evaluation, and learning-based auction for model trading.}
		\label{fig_usecase}
		\vspace{-0.2cm}
	\end{figure*}

	We consider pre-trained model sharing as model trading between source and target domains in a TL model trading market.  To achieve secure pre-trained model trading, a consortium blockchain named ``TL blockchain" is established on edge servers (e.g., RoadSide Units, RSU) acting as miners to run smart contracts for model information sharing and model trading of the pre-trained models \cite{9262056}. Specifically,
	\begin{itemize}
		\item \textit{Model Information Sharing Smart Contract}: Each model owner (e.g., a vehicle) first finalizes a training task (e.g., traffic sign recognition) in the source domain. The model owners generate metadata of their pre-trained models including pseudonyms of owners, model descriptions of the pre-trained models (e.g., training task, accuracy, model size, model usage, trading requirement, timestamp, and digital signature for information verification), and biding prices. The model owners then run the model information sharing smart contract to automatically upload the metadata to the TL blockchain miners, thus running a consensus algorithm, e.g., delegated proof-of-stake, to finish block verification and synchronization. After that, the pre-trained models can be recorded on the TL blockchain for model trading (as illustrated in Steps 1 and 2 in Fig.~\ref{fig_usecase}). 
		
		\item \textit{Model Trading Smart Contract}: Model buyers (e.g., vehicles) in the target domain obtain the latest block data from the TL blockchain. According to buyers' training tasks, e.g., pedestrian recognition, buyers search the metadata (e.g., past transaction and reputation records) on the block data and pick out potential model sellers (i.e., model owners in the source domain). Buyers also retrieve the reputation values of model sellers from the TL blockchain to filter malicious sellers with low reputations. Here, we utilize reputation value as a metric to indicate the reliability of model owners, and the reputation reflects the quality of pre-trained models of model owners (more details about reputation are given in Section~\ref{sec:framework}-B). Model buyers and sellers perform a learning-based auction scheme for model pricing and trading according to bidding prices and the reputation of the sellers (Section~\ref{sec:framework}-C). Then, model buyers obtain pre-trained models (i.e., knowledge) from the matched sellers and begin to fine-tune their target models. If the training performance of a target task is good, e.g., accuracy higher than a given threshold, the buyer will treat this trading record as a positive event and generate a good reputation rating (Steps 3, 4 and 5 in Fig.~\ref{fig_usecase})~\cite{2019kangFL}. After that, a model trading smart contract will be executed to upload the reputation and trading records to miners in the TL blockchain for block verification and recording (Step 6 in Fig.~\ref{fig_usecase}).
	\end{itemize}
	
	\subsection{Decentralized Reputation Management} 
	
	Similar to \cite{2019kangFL}, reputation is introduced to measure the trustworthiness and reliability of model sellers according to their historical behaviors. By the reputation, model buyers can select reliable and trusted model buyers with high-quality pre-trained models, thus avoiding the effects of low-quality models on target task training. Moreover, since reputation plays a significant role for TL, to remove the potential risks of centralized reputation calculation and management, we design a decentralized reputation calculation and management scheme utilizing advantages of TL and blockchain technologies.
	\begin{itemize}
		\item \textit{Decentralized reputation calculation:}
		For each traded pre-trained model from a seller, it will generate a performance rating of the pre-trained model using in the target tasks by model quality evaluation methods, e.g., sharply values \cite{parvez-chang-2021-evaluating}. If the pre-trained model leads to good performance on the target tasks, e.g., high accuracy, the model buyer treats this trading interaction with the model seller as a positive event, otherwise, the interaction is a negative event. Then, the direct reputation of buyer \textit{i} for seller \textit{j} is denoted as the ratio of the number of positive interaction events to the total number of interaction events in a time window. To achieve accurate and objective reputation calculation, we should consider both direct reputation and an overall recommended reputation from its friends, who have cooperated with the buyer before, for the same seller. Here, the overall recommended reputation is expressed by the weighted sum of the trustworthiness weights of buyers for their friends and the corresponding recommended reputation of the friend to the same seller. Both the trustworthiness weight and reputation value are in the range of [0, 1]. Moreover, the referenced reputation values from strangers are the average reputation value of a stranger to the seller. Finally, the integrated reputation value $r_{ij}$ of buyer \textit{i} to seller \textit{j} is the weighted sum of the direct reputation, the recommended reputation, and the referenced reputation values.
		
		\item \textit{Blockchain-based reputation management:}
		To achieve secure and decentralized reputation management, after model training,  the model buyers upload their integrated reputation values with the corresponding digital signatures as ``transactions" to the miners of the TL blockchain. With the help of an efficient consensus algorithm, the verified reputation data is stored in the TL blockchain for immutable, transparent, and reliable management. These reputation values will be recommended to and referenced by other model buyers.

	\end{itemize}
	
	\subsection{Deep Learning-based Auction Scheme for TL Model Trading Markets}
	
	To develop an efficient incentive mechanism, Auction Theory is regarded as a promising tool to maximize the income of model sellers while ensuring bidders' desired characteristics, such as individual rationality and incentive compatibility~\cite{niyato_luong_wang_han_2020}. However, traditional auction schemes cannot be applied to real-time optimal trading and pricing scenarios. We, therefore, develop a Deep Learning-based auction (DLA) scheme to make the allocating and pricing decisions for model sharing when demands of vehicles frequently change over time. {Particularly, an auctioneer parameterized by neural networks first performs monotone transformations of input buyers' values, as well as the reputation of sellers.} Based on the real-time reputations of sellers, the auctioneer then calculates the allocation rules, i.e., winning probabilities of bidders, and conditional payment rules for the TL model market. Lastly, the auctioneer is trained w.r.t. the loss function so that it can adjust the weights to optimize the expected outcome of this market. After the auctioneer is well-trained, it will be encoded in smart contracts and uploaded to the TL Blockchain, which is composed by RSUs in the 6G-enabled IoV.
	
	In DLA, the auctioneer knows the value distribution of bidders but not their actual valuation. The auctioneer inputs sellers' reputations and bidders' bids, and then outputs the pricing rules and the allocation rules (Steps 4.1 and 4.2 in Fig. \ref{fig_usecase}). The buyer with its valuation obtains its utility computed as valuation minus price. An auction is dominant strategy incentive compatible (DSIC)~\cite{dutting2019optimal}, if each buyer maximizes its utility by reporting truthfully regardless of its counterparts report. When each buyer receives a non-zero utility at the end of the auction, it is (ex-post) individually rational. The DSIC auction requires each buyer to report truthfully, and so the revenue on the valuation profile should be accurate. The DSIC auction that maximizes its revenue is considered to be the optimal DSIC auction design.
	
	To obtain a satisfactory solution for DSIC auction, DLA schemes are widely adopted. DLA utilizes multi-layer neural networks to encode auction mechanisms. In order to handle sequentially inputted bids for multiple items in the TL model markets, neural networks are used to process the current state and output the temporal results. In detail, reputations downloaded from the TL blockchain are input in the hidden layer. Thus, the proposed auction scheme is more reliable. Then, the temporal results are input to the softmax layer for allocation rules and the sigmoid layer for pricing rules. The goal of the DLA scheme is to find a solution to minimize the negative, expected revenue (Step 4.3 in Fig.~\ref{fig_usecase}), while satisfying incentive compatibility. Our primary goal is to ensure that the selected auction satisfies the requirement for incentive compatibility in the learning problem. Our method for measuring incentive compatibility is to calculate an ex-post regret for every bidder. Bidder is likely to regret their decision ex-post (Step 4.4 in Fig. \ref{fig_usecase}). In this context, the learning problem is to minimize the expected loss, i.e., the expected negated revenue s.t. sum ex-post regret for all bidders equals zero (Step 4.5 in Fig. \ref{fig_usecase}). Using the augmented Lagrangian method, we solve the constrained training problem over the space of neural autoworker parameters. Adding a quadratic penalty term for violating constraints, the Lagrangian function is defined for the optimization problem. Therefore, the DLA scheme can make secure and reliable optimal allocation and pricing decisions for the TL model market in real-time, hence preventing malicious vehicles from adversely affecting the performance of TL in 6G-enabled IoV. Furthermore, the computation complexity of the DLA scheme depends on the number of hidden layers and number of hidden neurons in hidden layers rather than the dimensions of the input layer and the output layer, i.e., size of markets. The complexity analysis of the DLA scheme with H hidden neurons is O((N+M)D+D$^2$) for N buyers and M sellers. Thus, the proposed scheme can be applied to large-scale TL model trading markets. In addition to addressing the TL model trading market, this secure and reliable framework is not limited by model trading scenarios. The framework is also capable of addressing other service trading problems requiring secure and reliable solutions.

	\subsection{Simulation Results} 
	Simulations are used to illustrate the effectiveness of our DL-based auction scheme. The proposed auction scheme is implemented with the same parameters in~\cite{dutting2019optimal}.
	TL model markets are simulated with different numbers of participants to analyze the performance of the DLA scheme. Here, TL model markets consist of 3 buyers and 3 sellers (3$\times$3) and 3 buyers and 5 sellers (3$\times$5). Figure 3 shows the simulation results for evaluation of the proposed secure and reliable TL framework. As shown in Fig. 3(a), the DLA and the DLA with low reputation sellers (DLA-LR) schemes all converge quickly to the solution. Compared with the second-price auction (SPA)~\cite{niyato_luong_wang_han_2020}, the DLA scheme can improve the revenue by about 30\%, while the DLA-LR can improve by about 10\%.
	Moreover, Fig. 3(b) shows the change of TL model accuracy affected by malicious sellers' attack strength and the permitted reputation of the TL market. We can observe that the increasing attack strength decreases the accuracy of target models. The higher reputation threshold of being sellers brings larger accuracy because fewer malicious models from low-reputation sellers are transferred to the buyers in the target domain.  Fig. 3(b) also illustrates that the reputation scheme can ensure reliable and secure TL by removing unreliable or malicious source model trainers. 
	\begin{figure}[t]
		\vspace{-0.28cm}
		\subfigure[Revenue vs. Epochs.]{
			\begin{minipage}[t]{0.5\linewidth}
				\includegraphics[width=1\linewidth]{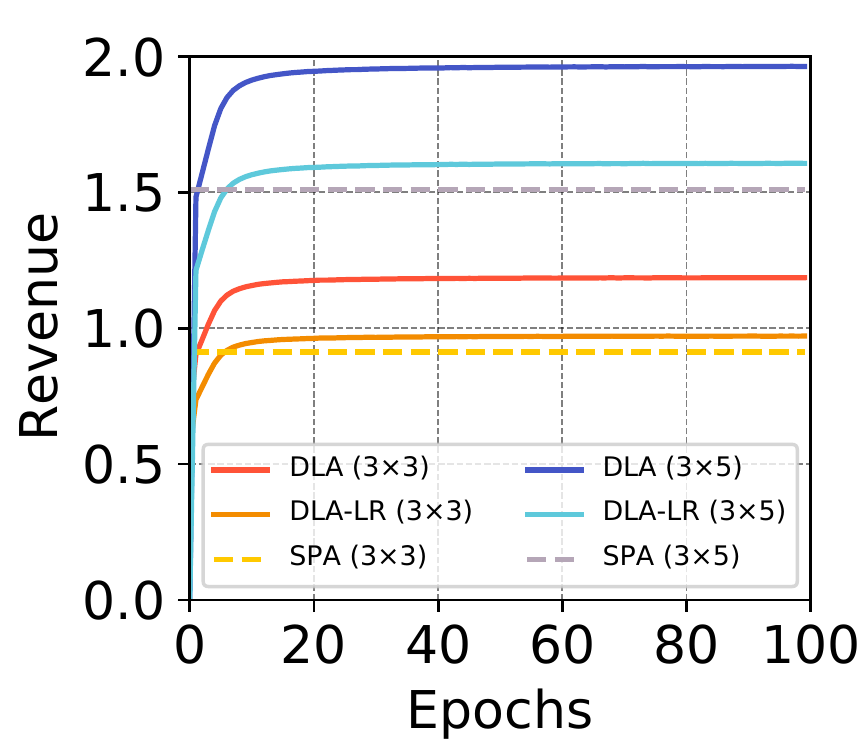}
			\end{minipage}%
		}%
		\subfigure[Reliability evaluation.]{
			\begin{minipage}[t]{0.5\linewidth}
				
				\includegraphics[width=1\linewidth]{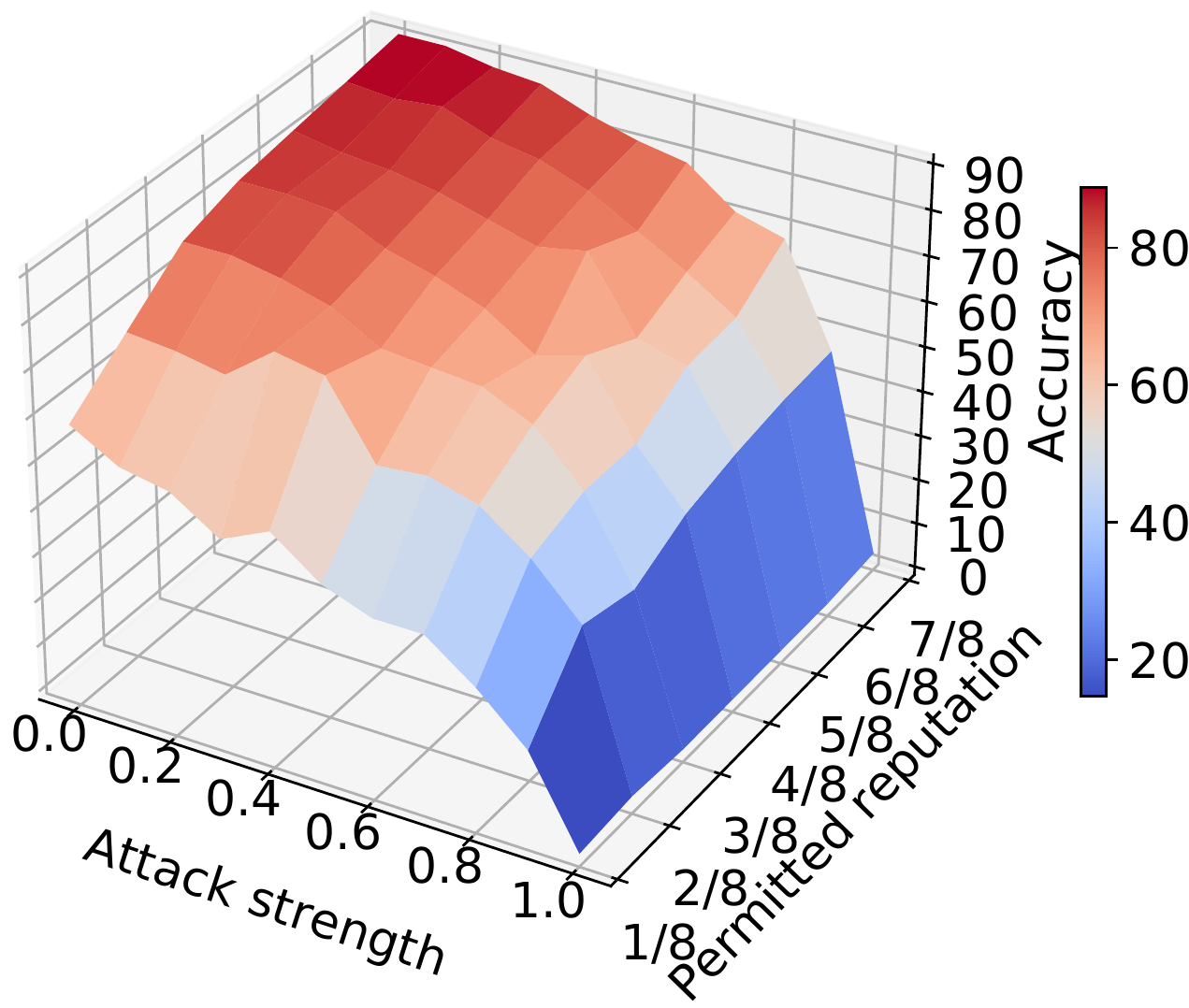}
			\end{minipage}%
		}%
		\centering
		\caption{Security and reliability evaluation of the proposed TL framework and DLA scheme, where the attack strength is the ratio of falsified labels of a training dataset of the sellers and the permitted reputation is the required reputation to allow sellers to participate.}
		\vspace{-0.28cm}
		\label{simulation}
	\end{figure}
	\section{Conclusions and Open Issues}
In this article, we propose a secure and reliable transfer learning framework for 6G-enabled IoV with blockchain empowered model trading market. In the model trading market, we first introduce reputation as the metric to evaluate the reliability of model sellers in the source domain. With the help of reputation, we then design the machine learning-based auction scheme considering sellers' reputations to achieve optimal DSIC auction in model trading pricing for reliable model trading in TL. The simulation results show that the proposed framework and scheme can provide secure and reliable transfer learning services with well-design incentives. 

There are several possible directions that are worth being studied: 1) Since vehicles in 6G networks have limited power and resources. It still remains to be an open issue on how to design light-weight and resource-efficient TL algorithms for large-scale deployment on these vehicles by combining advanced machine learning techniques, such as sparse representation in neural networks, and pruning algorithms to dwindle TL models. 2) Since  model quality valuation methods directly affect  the reputation calculation, there is one more open issue on how to design more accurate and efficient model quality valuation methods  for 6G-enable IoV to improve the accuracy and objectivity  of reputation calculation, thus enhancing the detection performance of unreliable model sellers. 3) It is also an open issue on how to improve the scalability of consortium blockchain for efficient reputation management in TL-based 6G-enabled IoV. Sharding and cross-chain technologies are promising solutions to establish scalable and efficient blockchain systems for 6G-enabled IoV. 4) Considering the high overhead of a large number of vehicles that may join in the transfer learning for 6G-enabled IoV, efficient schemes for optimizing the number of training model in target domains are worth investigating to balance the learning performance and resource cost.

	\label{sec:Sum}

	\bibliography{TF_for_6G_IoV}
\end{document}